\documentclass{article}

\usepackage[utf8]{inputenc} 
\usepackage[T1]{fontenc}    
\usepackage{hyperref}       
\usepackage{url}            
\usepackage{booktabs}       
\usepackage{amsfonts}       
\usepackage{nicefrac}       
\usepackage{microtype}      
\usepackage{xcolor}         

\usepackage{natbib}

\usepackage{todonotes}

\usepackage{siunitx}

\title{Modeling and Optimizing Laser-Induced Graphene}

\author{%
    Lars Kotthoff, Sourin Dey, Vivek Jain, Alexander Tyrrell,\\Hud Wahab, Patrick Johnson\\
  Center for Artificially Intelligent Manufacturing\\
  University of Wyoming\\
}

\begin{document}

\maketitle

\begin{abstract}
A lot of technological advances depend on next-generation materials, such as graphene, which enables a raft of new applications, for example better electronics. Manufacturing such materials is often difficult; in particular, producing graphene at scale is an open problem. We provide a series of datasets that describe the optimization of the production of laser-induced graphene, an established manufacturing method that has shown great promise. We pose three challenges based on the datasets we provide -- modeling the behavior of laser-induced graphene production with respect to parameters of the production process, transferring models and knowledge between different precursor materials, and optimizing the outcome of the transformation over the space of possible production parameters. We present illustrative results, along with the code used to generate them, as a starting point for interested users. The data we provide represents an important real-world application of machine learning; to the best of our knowledge, no similar datasets are available.
\end{abstract}

\section{Introduction}

Graphene is a two-dimensional honeycomb layer of carbon atoms with extraordinary
properties, for example relative strength higher than any other material, high
conductivity of electricity and heat, and near transparency. It has many
promising applications, such as next-generation semiconductors, flexible
electronics, and smart windows, to name but a few
examples \citep{ferrari_science_2015}. There already exist a number of
commercially available products made from or with graphene, and the size of the
global market is currently about US-\$100 million, with significant growth
forecast. However, the reliable and large-scale production of graphene is a
difficult problem that researchers have been tackling over the past decades.

One method of producing graphene is to convert natural sources of carbon, e.g.\
graphite, coal, and biochar, into graphene oxide, which is soluble in water.
Such solutions can be used as graphene oxide inks and be printed directly onto
substrates as thin films, similar to how ink-jet printers deposit ink on paper.
Irradiating this precursor material with a laser heats and anneals the graphene
oxide selectively to reduce the oxygen, ultimately converting it into pure
graphene. Similar results can be achieved by irradiating commercial polymer
films, eliminating the need to manufacture and deposit graphene oxide, which is
time-consuming in itself, or indeed any carbon precursor material
\citep{chyan_laser_2018}. The reduction of such precursor materials into
graphene allows for the rapid and chemical-free manufacturing of advanced
devices such as electronic sensors \citep{luo_direct_2016}, fuel cells
\citep{ye_situ_2015}, supercapacitors
\citep{lin_laser-induced_2014,el-kady_scalable_2013}, and solar cells
\citep{sygletou_laser_2016}. The interested reader is referred to a recent
survey on laser-induced graphene for more information
\citep{wang_laser-induced_2018}. This process is also referred to as
laser-reduced graphene in the literature \citep{zhengfen_laser_2018}.

One of the advantages of the targeted irradiation of the precursor material is
that it allows to easily create patterns in solid substrates without
pre-patterned masks in only a few minutes. While graphene is electrically
conductive, graphene oxide and polymers are not -- patterns of graphene in an
insulating material can form electric circuits. The laser irradiation process
enables the scalable and cost-efficient fabrication of miniaturized electronic
devices in a single process, rather than manufacturing the graphene separately
and then patterning it onto a carrier material. This process also ensures that
only the amount of material that is actually needed is produced, similar to
other advanced manufacturing processes like 3D printing.

The challenge in irradiating the precursor material is determining the best
laser parameters and reaction environment. First-principles
knowledge does not allow to derive the optimal conditions and the effectiveness
of different irradiation conditions varies across different precursor materials.
A recent study emphasizes the effect the irradiation parameters have on the
quality of the produced graphene and the need to optimize these parameters to
achieve good results in practice \citep{wan_tuning_2019}. Even with just a few
parameters, for example the power applied to the laser and the duration for
irradiating a particular spot, the space of possibilities is too large to
explore exhaustively. There are complex interactions between parameters, and
evaluating a particular parameter configuration involves running an experiment that
requires a skilled operator and precursor material of sufficient quality.
Exploring the space of experimental parameters efficiently is crucial to the
success of laser-induced graphene in practice. In many cases, this optimization
is guided by human biases -- an area ripe for the application of machine
learning.

We have applied Bayesian optimization to the automated production of
laser-induced graphene, improving the quality of the produced graphene
significantly compared to results achieved in the literature
\citep{wahab_machine-learning-assisted_2020}. In this paper, we present a series
of datasets obtained in the process for the community to build on. To the best
of our knowledge, there are no similar datasets. In particular, the data we make
freely available represents an important and challenging application of machine
learning in a rapidly-growing industry. Beyond graphene, materials science in
general is an increasingly prominent application area of machine learning. We
outline possible uses for the data, along with illustrative results. All data,
code, and results are available at
\url{https://github.com/aim-uwyo/lig-model-opt}.

\section{Methodology}

The graphene oxide samples used for the data we present here were prepared from
graphite using the improved Hummers' method \citep{hummers}. Powdered samples,
ground and sieved to \SI{20}{\micro\meter}, were mixed in concentrated $H_2SO_4$
and $H_3PO_4$ and placed in an ice bath. $KMnO_4$ was added at a mixture
temperature of \SI{35}{\celsius} and increased to \SI{98}{\celsius} before
termination with ultrapure water (Millipore) and $H_2O_2$. The filtrate was then
washed with $HCl$ and subsequently with water repeatedly until a pH-level of
about 6.5 was obtained. The GO inks were produced using \SI{25}{\milli\gram} of
the freeze-dried GO powder, which was diluted in \SI{100}{\ml} deionized water
and ultrasonicated with a cooling system. After the sample was centrifuged, the
remaining supernatant was repeatedly diluted and ultrasonicated until a
\SI{200}{\ml} dilution was obtained. The GO inks were spray-coated onto a
\SI{1}{\cm}$\times$\SI{1}{\cm} quartz or polyimide substrate (Kapton HN
\SI{125}{\micro\meter}, Dupont) in multiple passes until a thickness of
\SI{1}{\micro\meter} was achieved, verified with an optical profilometer.

\begin{figure}[htb]
    \begin{center}
        \includegraphics[width=.8\textwidth]{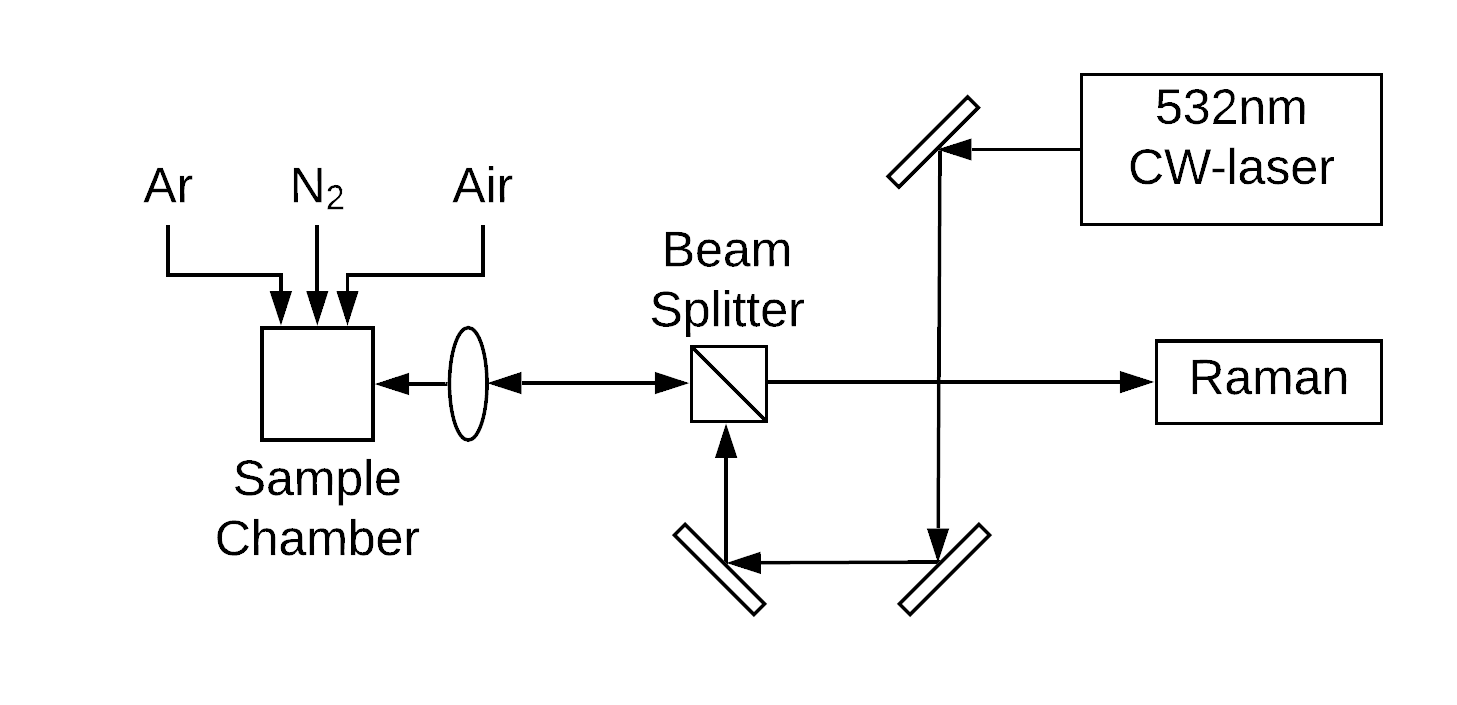}
        \caption{Experimental setup for patterning and measuring laser-induced
        graphene. The unlabeled rectangles represent mirrors to reflect the
        laser beam, the ellipse a lens to focus it.}
        \label{fig:setup}
    \end{center}
\end{figure}

Laser-induced graphene (LIG) spots were patterned by reducing GO films deposited
on quartz and polyimide, and by carbonization of polyimides directly. We denote
GO on quartz, GO on polyimide and polyimide as samples GOQ, GOPI and PI,
respectively. The patterning setup is shown in Figure~\ref{fig:setup}. The
deposited GO films were placed in a sample chamber which allows patterning in
air, argon, or nitrogen environments with pressures up to 1000 psi. LIG patterns
were irradiated using a \SI{532}{\nm} diode-pumped solid-state continuous-wave
laser. The laser beam was focused with a 50x microscope lens to a spot size of
\SI{20}{\micro\meter} on the sample surface. Irradiated beam spots were
positioned sufficiently far apart from each other to ensure pristine precursor
material for each experiment. The sample area is about \SI{1}{\square\cm},
allowing approximately 256, 25, and 25 patterns for samples GOQ, GOPI, and PI,
respectively. Taking into account sample preparation and repeated measurements
to account for experimental errors and ensure reproducibility, we set our
experimental budget to 70 for all types of samples.

\begin{figure}[htb]
    \begin{center}
        \includegraphics[width=.5\textwidth]{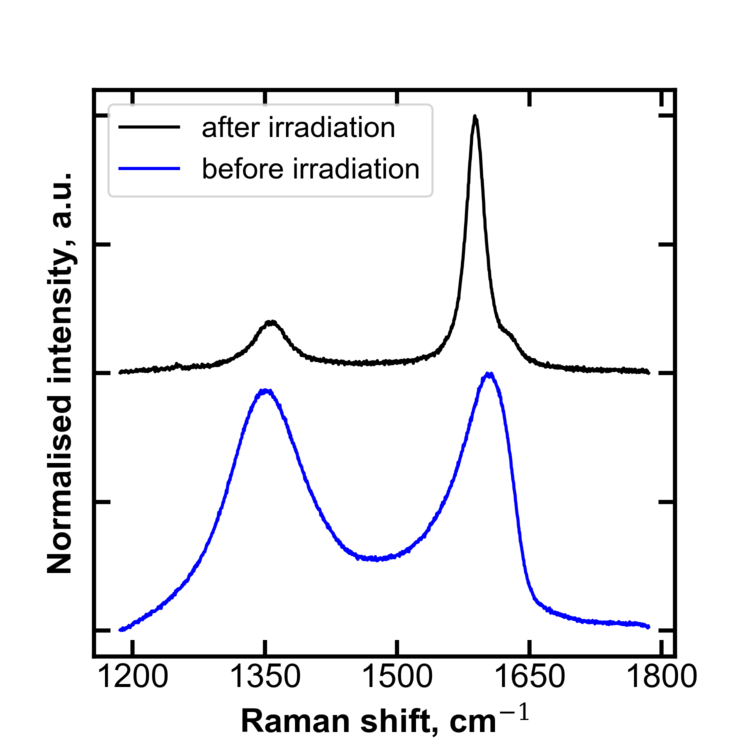}
        \caption{Raman spectra showing D (left peak) and G (right peak) bands of
            graphene oxide before (bottom) and after (top) laser irradiation.
            The ratio we optimize in this paper is calculated from the area
        under the peaks. The intensity is shown in arbitrary units (a.u.).}
        \label{fig:raman}
    \end{center}
\end{figure}

Raman spectroscopy is a common technique for determining the quality of
laser-induced graphene by observing how laser photons scatter after they
interact with the vibrating molecules in the sample probe. The intensities of
the characteristic D and G bands in the Raman spectra can be used to judge to
what extent the precursor material has been reduced to graphene, i.e.\ the
quality of the resulting material. The D and G bands result from the defects and
in-plane vibrations of sp$^2$ carbon atoms, respectively. In particular, the
degree of reduction of the precursor material to graphene, and thus the
conductivity of the irradiated area, can be quantified through the ratio of the
intensities of the G and D bands -- the larger this ratio, the more the
precursor material has been reduced. Figure~\ref{fig:raman} shows an example.

We filtered the backscattered laser beam through a long-pass filter after
irradiation to perform Raman spectroscopy. Using the same laser source for
patterning and spectroscopy, we are able to characterize the identical spot
in-situ. The Raman data for each spot were averaged over $10$ measurements with
a collection time of \SI{3}{\second} at laser power $<$\SI{10}{\mW} for each
measurement. The Raman spectra were post-processed with a linear background
subtraction to $0$ and normalization of the maximum peak to $1$. The G- and
D-bands were fitted using Lorentzian functions and the ratio of their
intensities computed as the ratio of the areas under the fitted functions. The
G/D ratios indicate the degree of reduction of GO to graphene. This measure can
be used as a proxy for electric conductivity, which determines the suitability
of the produced material for advanced electronics. More information on the
experimental setup can be found in \citep{wahab_machine-learning-assisted_2020}.

\subsection{Parameter Space}

We consider the following four parameters of the experimental conditions that
control the irradiation process.
\begin{itemize}
    \item The power applied to the laser used to irradiate the sample. We
        consider a power range of \SIrange{10}{5550}{\milli\watt}.
    \item The duration a particular spot was irradiated by the laser. We vary
        this parameter from \SIrange{500}{20000}{\milli\second}.
    \item The pressure in the reaction chamber. The values for this parameter
        range from \SIrange{0}{1000}{psi}.
    \item The gas in the reaction chamber. Possible values for this parameter
        are argon, nitrogen, and air.
\end{itemize}

These parameters give rise to a large space of possible combinations that is
infeasible to explore exhaustively. The cost of gathering data is high
-- running experiments is time-consuming and requires precursor materials to be
available. In contrast to big-data approaches, we need techniques that work with
small amounts of data, such as the Bayesian optimization approach we applied to
gather the data we present here.

\subsection{Bayesian Optimization}

Bayesian model-based optimization techniques (MBO) are used in many areas of
machine learning and AI and beyond to automatically optimize outcomes across
large parameter spaces. They usually proceed in an iterative fashion -- they
predict the configuration to evaluate, and the result of this evaluation informs
the predictions for the configuration to evaluate next. At the heart of these
techniques are so-called surrogate models, which approximate and model the
process whose parameters are to be tuned. This underlying process is expensive
to evaluate, i.e.\ it is infeasible to exhaustively explore the parameter space
and we are interested in keeping the number of evaluations as small as possible.
The approximate surrogate model on the other hand is cheap to evaluate and
allows for a targeted exploration of the parameter space, identifying promising
configurations that available resources for evaluations of the underlying
process should be directed towards.

Surrogate models are induced using machine learning, taking an increasing amount
of ground-truth data into account between subsequent iterations.
State-of-the-art MBO approaches often use Gaussian Processes or random forests
to induce surrogate models, depending on the nature of the parameter space. MBO
is a mature approach that has been used in many applications over decades, for
example in automated machine learning
\citep{feurer_efficient_2015,kotthoff_auto-weka_2017}. The interested reader is
referred to the paper that formalized the approach \citep{jones_efficient_1998}
for more information.

There are many implementations of MBO; we use the mlr\-MBO package \citep{mlrMBO}
to model the parameter space, build the surrogate models (with the mlr
package \citep{bischl_mlr_2016}), and determine the most promising configuration
for the next evaluation of the underlying process. In particular, we use the
default random forest surrogate model for parameter spaces that contain
non-continuous parameters (the gas in the reaction chamber) and expected
improvement as our acquisition function. In each iteration of the optimization
process, the next configuration to evaluate is proposed by mlrMBO. This
configuration is set automatically by the experimental setup, which proceeds
with running the experiment and evaluating its result. The evaluated parameter
configuration and the resulting G to D ratio of the irradiated spot is added to
the data used to train the surrogate model for the next iteration. We present
the datasets obtained when the process ends.

For the initial surrogate model, we evaluated 20 parameter configurations that
were randomly sampled from the entire parameter space. We then performed 50
iterations of our model-based optimization approach, for the total 70
evaluations we can perform on a single sample. For each of the three
investigated materials GOQ, GOPI, and PI, we ran three experimental campaigns
for a total of nine experimental campaigns and 630 patterned spots, which
represents several weeks of experimental effort, in addition to the effort of
preparing the samples.

\section{Machine Learning Datasets from Materials Science}

We present three datasets from the above application. All datasets describe the
transformation from a precursor material into graphene through laser
irradiation, where the quality of the result depends on the parameters of the
laser and reaction environment. The difference between the different datasets is
that they were obtained with different precursor materials, as described above.
Each dataset consists of three experimental campaigns each, for a total of 210
data points per dataset. Metadata is provided for each datum indicating which
experimental campaign it belongs to and whether it was part of the initial,
randomly sampled data, or proposed by the Bayesian optimization process.

We envision three different areas of machine learning where this data will be
useful; we describe each below with illustrative results and the code that was
used to produce them. Our illustrative results are intended to show what
performance can be achieved to give prospective users a starting point; we do
not make any claims with respect to the optimality of our results.

The data, scripts to produce the illustrative results we describe below, and the
figures themselves are available at
\url{https://github.com/aim-uwyo/lig-model-opt} under the permissive 3-clause
BSD license. No ethical issues arose in gathering the data, but we caution that
they could potentially be used in unethical applications, for example to produce
advanced electronics for weapons systems. We do not condone or encourage such
applications.

\subsection{Modeling Laser-Induced Graphene}

The datasets we provide can be used in straightforward manner to predict the
quality of the transformation of the precursor material into graphene, given the
experimental parameters. We ran illustrative experiments with the mlr3 machine
learning toolkit~\citep{lang_mlr3_2019}. The code necessary to reproduce our
results is provided in supplementary material, together with the data. In
addition, we ran experiments with the auto-sklearn~\citep{feurer_efficient_2015}
automated machine learning toolkit, running it with a time limit of one hour.

\begin{figure}[!htb]
    \begin{center}
        \includegraphics[width=\textwidth]{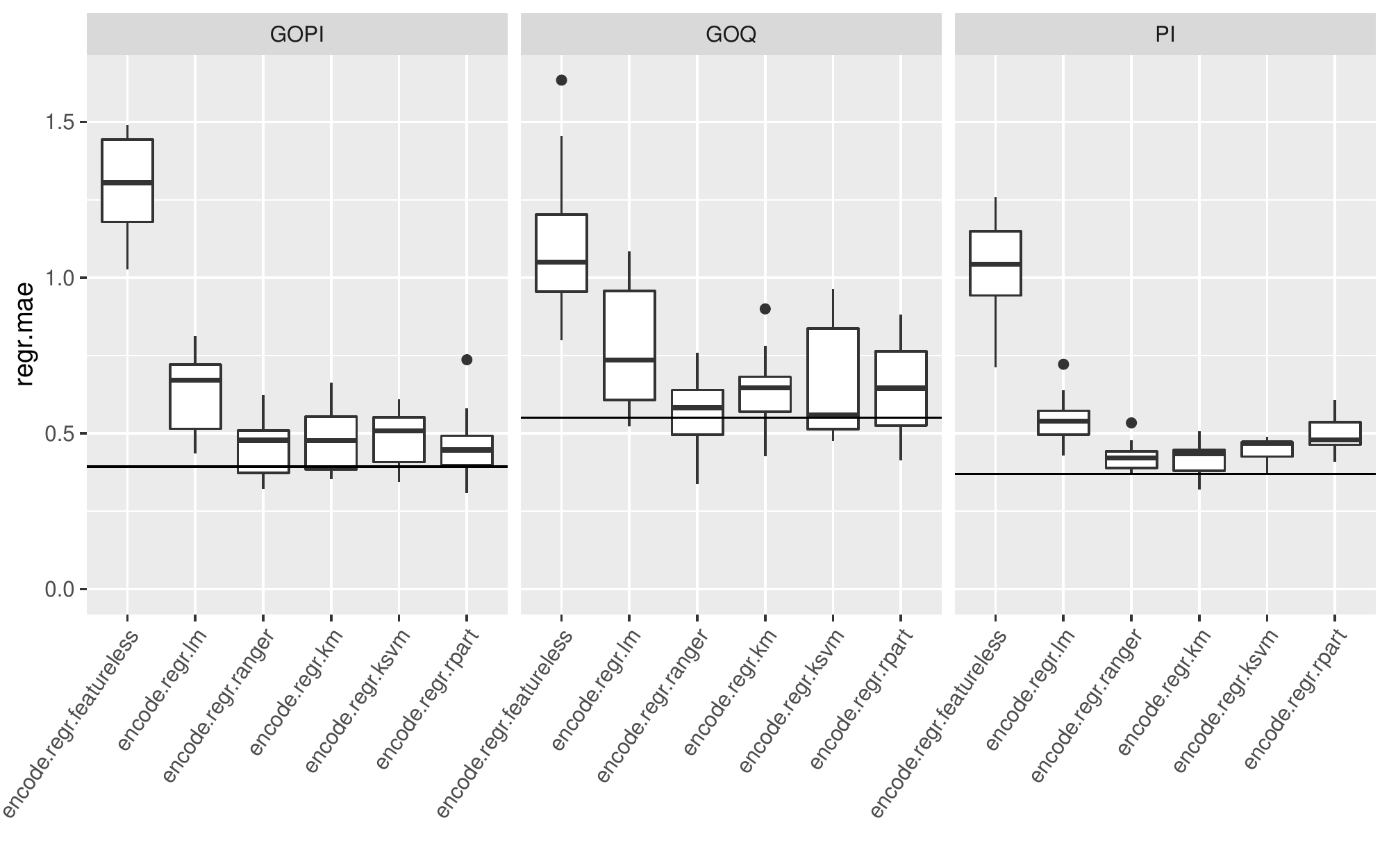}
        \caption{Illustrative results for modeling the transformation of the
            precursor material into graphene for different machine learning
            approaches. We show the mean absolute error over 10 cross-validation
            folds for (from left to right) a dummy featureless learner, a simple
            linear model (lm), a random forest (ranger), a Gaussian Process
            (km), a support vector machine (ksvm), and a regression tree
            (rpart). The featureless learner simply predicts the mean value of
            the training set. The horizontal lines denote the performance of
            auto-sklearn on each of the datasets.}
        \label{fig:modeling}
    \end{center}
\end{figure}

We present illustrative results in Figure~\ref{fig:modeling}. Even simple approaches,
such as linear models, already achieve much better performance than the baseline
featureless learner. More sophisticated approaches, such as the Gaussian
Processes and random forests that are ubiquitous surrogate models in Bayesian
optimization, do not further performance much. The same is true for much more
sophisticated automated machine learning approaches, especially on the GOQ
dataset.

Results are best for the GOPI and PI datasets in terms of improvement over the
baseline, with GOQ showing a smaller gap. We believe that model performance can
be increased further; in particular, automated machine learning has been able to
improve performance only slightly here.

We note that, in contrast to most machine learning datasets, the data we present
here is not identically and independently distributed, as it has been obtained
as part of Bayesian optimization runs, where a data point depends on the
previous ones. This violates the basic assumption underlying most machine
learning approaches. In practice, building surrogate models from non-i.i.d.\
data appears to work fine, as good results from applying Bayesian optimization,
including ours, show. Nevertheless, the implications of using non-i.i.d.\ data
in this context are understudied, and our data provides and opportunity to do
so. Each point in the raw data has metadata denoting whether the point was
obtained as part of the initial, random and i.i.d., data or evaluated in a
Bayesian optimization iteration.

\subsection{Transfer Learning}

Three datasets from very similar but different setups also provide the
opportunity to explore to what extent knowledge acquired from one dataset can be
transferred to another. While the process is the same, the precursor materials
are different and react differently to the same experimental conditions. There
are latent features that encode the properties specific to each precursor
material that machine learning may be able to extract.

\begin{figure}[!htb]
    \begin{center}
        \includegraphics[width=\textwidth]{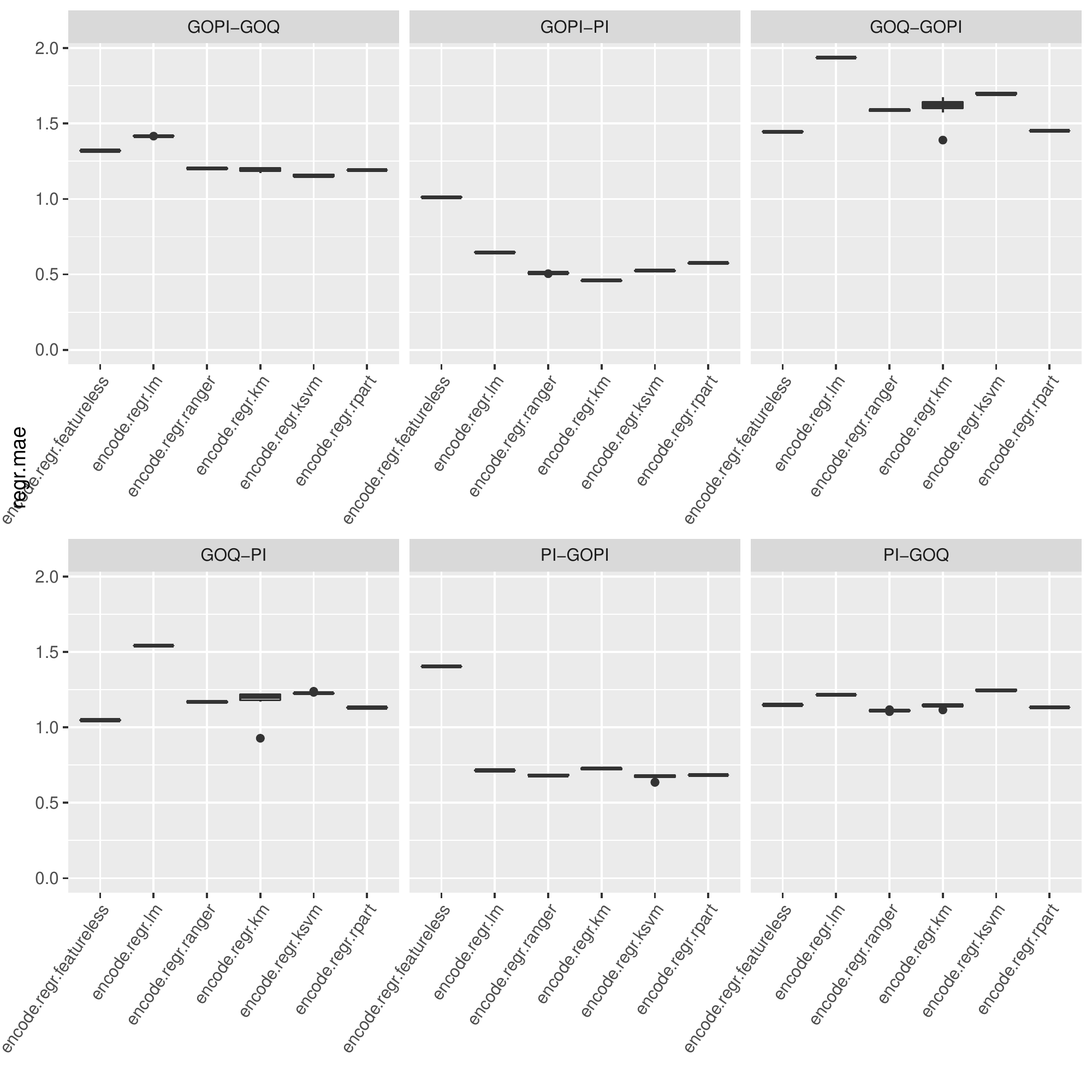}
        \caption{Illustrative results for transferring models from one precursor
            material to another. We show the mean absolute error for the same
            learners as above, including the dummy featureless learner. The
            first dataset in the title of a plot denotes the training set, while
            the latter denotes the test set. We randomly sample 80\% of the
            respective datasets for training and test, repeated 10 times.}
        \label{fig:transfer1}
    \end{center}
\end{figure}

We ran illustrative experiments, again with the mlr3 machine learning toolkit.
Figure~\ref{fig:transfer1} shows illustrative results from the evaluation of a model
learned on one dataset on another. It is immediately clear that the two
precursor materials based on polyimide, GOPI and PI, behave very similarly --
models trained on one precursor material transfer with good performance to the
other, although not as good as for models trained and evaluated on the same
dataset (two middle panels in the figure). For GOQ, transferred models (both
from and to this precursor material) do not show good performance compared to
the baseline model, indicating that the precursor materials are sufficiently
different that a direct transfer is infeasible.

\begin{figure}[!htb]
    \begin{center}
        \includegraphics[width=\textwidth]{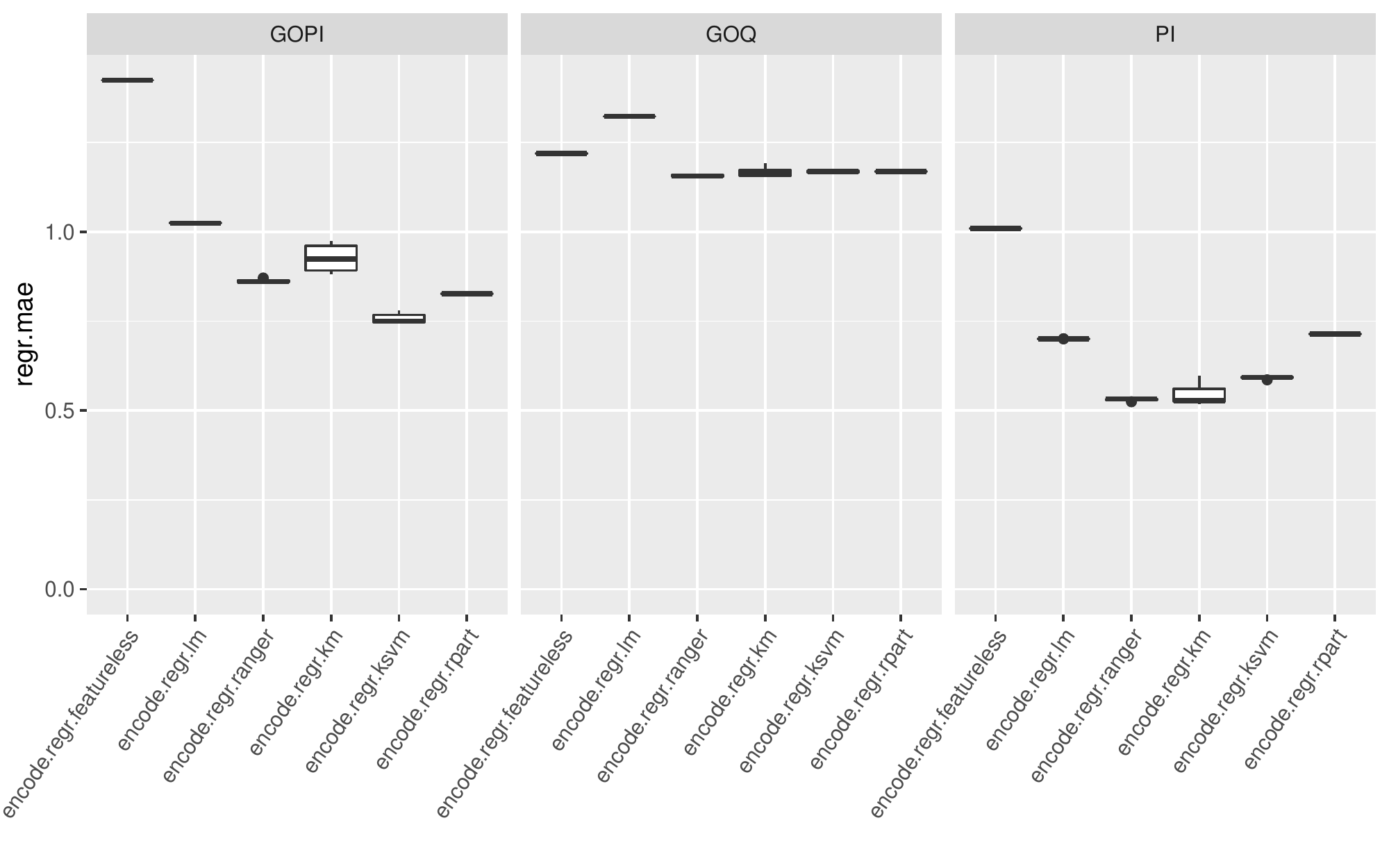}
        \caption{Illustrative results for transferring models trained on two
            precursor materials to the other. We show the mean absolute error
            for the same learners as above, including the dummy featureless
            learner. The title of the dataset denotes the one that the
            performance of the models learned on the other two was tested on. We
            randomly sample 80\% of the respective datasets for training and test,
            repeated 10 times.}
        \label{fig:transfer2}
    \end{center}
\end{figure}

We further explore the performance of transferred models in
Figure~\ref{fig:transfer2}, this time by training models on the combination of
datasets of two precursor materials and evaluating their performance on the
dataset of the third precursor material. We again see that the two precursor
based on polyimide are quite similar, while GOQ is different and transferred
models do not exhibit good performance.

We provide two datasets that are quite similar, GOPI and PI, and one that is
quite different from the others, GOQ. This allows to create ``easy'' and
``hard'' transfer learning scenarios.

\subsection{Bayesian Optimization}

The datasets we provide can also be used for Bayesian optimization, which is how
the data was obtained to start with. In the end, we are interested in the best
conversion of precursor material into graphene and better ways of obtaining the
experimental parameters for that. An interesting aspect of the data we provide
is that it comes from a real-world application with a good motivation for
applying a sample-efficient optimization method, as obtaining new data points is
extremely expensive.

\begin{figure}[!htb]
    \begin{center}
        \includegraphics[width=\textwidth]{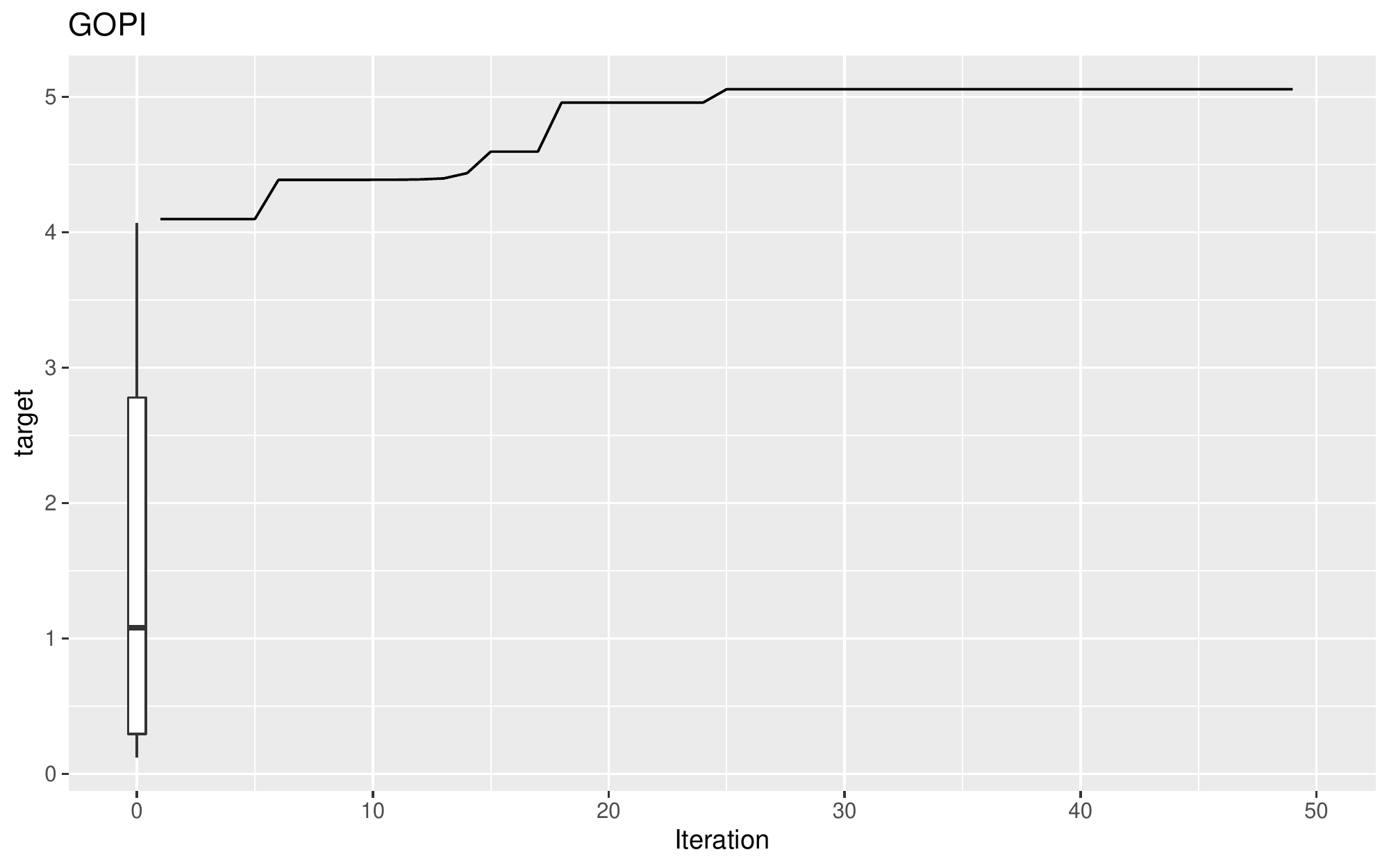}
        \caption{Illustrative results for Bayesian optimizations on simulators
            trained on entire datasets, here for the GOPI precursor material.
            The boxplot at iteration zero shows the distribution of the initial,
            randomly sampled data, while the line shows the cumulative best
            achieved transformation from the precursor material into graphene
            (measured by the G/D ratio) over the iteration number of the Bayesian
            optimization.}
        \label{fig:mbosim}
    \end{center}
\end{figure}

We provide simulators based on surrogate models built on entire datasets to
facilitate Bayesian optimization. Simulators and illustrative experiments are
based on the mlr~\citep{bischl_mlr_2016} and mlrMBO~\citep{mlrMBO} toolkits; the same we
used in the publication related to our datasets.\footnote{While the mlr toolkit
has been superseded by mlr3, the corresponding successor to mlrMBO is not
available yet. The underlying machine learning algorithms are the same.}
Figure~\ref{fig:mbosim} shows illustrative results. We show only results for the GOPI
precursor material for space reasons; results for the other precursor materials
are qualitatively similar and available at
\url{https://github.com/aim-uwyo/lig-model-opt}. Interested users can easily
plug in their own approach and evaluate how efficiently and effectively it
explores the optimization landscape provided by the surrogate models.

There are multiple ways our data can be used to improve the Bayesian
optimization process. Better surrogate models will enable better optimization,
and can be explored independently. Similarly, being able to transfer knowledge
from other Bayesian optimization runs, for example on different precursor
materials, will improve performance. Both of these challenges can be pursued
with the datasets we provide, in addition to methodological improvements to
Bayesian optimization.

Explaining black-box machine learning models is becoming increasingly important,
especially for real-world applications like the one we present here. On one
hand, being able to understand a model increases trust in it, while on the other
hand a machine-learned model may have acquired insights that are unknown to
humans and may advance our scientific understanding of the optimized process. We
explore some such methods in~\citep{wahab_machine-learning-assisted_2020}, but
there is scope for further exploration and explanation of the surrogate models.

We note that the simulators we provide can be used to evaluate different
optimization methods, such as genetic algorithms or Hyperband, equally as well.
We focus on Bayesian optimization here as this is the methodology we applied for
the application itself, but what we provide is not limited to that.

\section{Conclusions and Outlook}

We have presented three datasets drawn from a real-world application of machine
learning; the production of laser-induced graphene. The data are accompanied by
metadata and example code that demonstrates possible uses. To the best of our
knowledge, it is the first series of datasets from materials science with the
presented level of comprehensiveness, and we hope that it will facilitate and
inspire more applications of machine learning in this area and beyond.

Gathering the data we make available took significant effort, from preparing the
samples, running the experiments, to post-processing the raw experimental data.
This is common in materials science, where gathering data often involves
synthesizing a material or performing an experiment that leads to its
transformation or destruction. For this reason, big data methods are not
applicable here, or may only be applied with difficulty. We hope that by making
our data available, we will stimulate research on small data and
sample-efficient methods.

The code used to obtain the illustrative results we present here is available as
part of the datasets, and all results are fully reproducible. This provides an
easy starting point for interested users. We place no restrictions on the use of
the code and data we make available, but discourage unethical uses.

\subsection*{Acknowledgments}

We are supported by the University of Wyoming's College of Engineering and
Applied Sciences' Engineering Initiative, the School of Energy Resources at the
University of Wyoming, the Wyoming NASA Space Grant Consortium, and NASA EPSCoR.
SD and LK are supported by NSF award \#1813537. The sponsors had no involvement
in the creation of the datasets or this manuscript.

\bibliographystyle{plainnat}
\bibliography{\jobname}

\begin{thebibliography}{18}
\providecommand{\natexlab}[1]{#1}
\providecommand{\url}[1]{\texttt{#1}}
\expandafter\ifx\csname urlstyle\endcsname\relax
  \providecommand{\doi}[1]{doi: #1}\else
  \providecommand{\doi}{doi: \begingroup \urlstyle{rm}\Url}\fi

\bibitem[Bischl et~al.(2016)Bischl, Lang, Kotthoff, Schiffner, Richter,
  Studerus, Casalicchio, and Jones]{bischl_mlr_2016}
Bernd Bischl, Michel Lang, Lars Kotthoff, Julia Schiffner, Jakob Richter, Erich
  Studerus, Giuseppe Casalicchio, and Zachary~M. Jones.
\newblock mlr: {Machine} {Learning} in {R}.
\newblock \emph{Journal of Machine Learning Research}, 17\penalty0
  (170):\penalty0 1--5, 2016.
\newblock URL \url{http://jmlr.org/papers/v17/15-066.html}.

\bibitem[Bischl et~al.(2017)Bischl, Richter, Bossek, Horn, Thomas, and
  Lang]{mlrMBO}
Bernd Bischl, Jakob Richter, Jakob Bossek, Daniel Horn, Janek Thomas, and
  Michel Lang.
\newblock {{mlrMBO}}: {{A Modular Framework}} for {{Model}}-{{Based
  Optimization}} of {{Expensive Black}}-{{Box Functions}}.
\newblock \emph{arXiv:1703.03373}, 2017.
\newblock URL \url{http://arxiv.org/abs/1703.03373}.

\bibitem[Chyan et~al.(2018)Chyan, Ye, Li, Singh, Arnusch, and
  Tour]{chyan_laser_2018}
Yieu Chyan, Ruquan Ye, Yilun Li, Swatantra~Pratap Singh, Christopher~J.
  Arnusch, and James~M. Tour.
\newblock Laser-induced graphene by multiple lasing: Toward electronics on
  cloth, paper, and food.
\newblock \emph{ACS Nano}, 12\penalty0 (3):\penalty0 2176--2183, 2018.
\newblock \doi{10.1021/acsnano.7b08539}.

\bibitem[El-Kady and Kaner(2013)]{el-kady_scalable_2013}
Maher~F. El-Kady and Richard~B. Kaner.
\newblock Scalable fabrication of high-power graphene micro-supercapacitors for
  flexible and on-chip energy storage.
\newblock \emph{Nature Communications}, 4:\penalty0 1475, February 2013.
\newblock URL \url{https://doi.org/10.1038/ncomms2446}.

\bibitem[Ferrari et~al.(2015)Ferrari, Bonaccorso, Fal'ko, Novoselov, Roche,
  Bøggild, Borini, Koppens, Palermo, Pugno, Garrido, Sordan, Bianco,
  Ballerini, Prato, Lidorikis, Kivioja, Marinelli, Ryhänen, Morpurgo, Coleman,
  Nicolosi, Colombo, Fert, Garcia-Hernandez, Bachtold, Schneider, Guinea,
  Dekker, Barbone, Sun, Galiotis, Grigorenko, Konstantatos, Kis, Katsnelson,
  Vandersypen, Loiseau, Morandi, Neumaier, Treossi, Pellegrini, Polini,
  Tredicucci, Williams, Hee~Hong, Ahn, Min~Kim, Zirath, van Wees, van~der Zant,
  Occhipinti, Di~Matteo, Kinloch, Seyller, Quesnel, Feng, Teo, Rupesinghe,
  Hakonen, Neil, Tannock, Löfwander, and Kinaret]{ferrari_science_2015}
Andrea~C. Ferrari, Francesco Bonaccorso, Vladimir Fal'ko, Konstantin~S.
  Novoselov, Stephan Roche, Peter Bøggild, Stefano Borini, Frank H.~L.
  Koppens, Vincenzo Palermo, Nicola Pugno, José~A. Garrido, Roman Sordan,
  Alberto Bianco, Laura Ballerini, Maurizio Prato, Elefterios Lidorikis, Jani
  Kivioja, Claudio Marinelli, Tapani Ryhänen, Alberto Morpurgo, Jonathan~N.
  Coleman, Valeria Nicolosi, Luigi Colombo, Albert Fert, Mar Garcia-Hernandez,
  Adrian Bachtold, Grégory~F. Schneider, Francisco Guinea, Cees Dekker, Matteo
  Barbone, Zhipei Sun, Costas Galiotis, Alexander~N. Grigorenko, Gerasimos
  Konstantatos, Andras Kis, Mikhail Katsnelson, Lieven Vandersypen, Annick
  Loiseau, Vittorio Morandi, Daniel Neumaier, Emanuele Treossi, Vittorio
  Pellegrini, Marco Polini, Alessandro Tredicucci, Gareth~M. Williams, Byung
  Hee~Hong, Jong-Hyun Ahn, Jong Min~Kim, Herbert Zirath, Bart~J. van Wees,
  Herre van~der Zant, Luigi Occhipinti, Andrea Di~Matteo, Ian~A. Kinloch,
  Thomas Seyller, Etienne Quesnel, Xinliang Feng, Ken Teo, Nalin Rupesinghe,
  Pertti Hakonen, Simon R.~T. Neil, Quentin Tannock, Tomas Löfwander, and Jari
  Kinaret.
\newblock Science and technology roadmap for graphene, related two-dimensional
  crystals, and hybrid systems.
\newblock \emph{Nanoscale}, 7\penalty0 (11):\penalty0 4598--4810, 2015.
\newblock \doi{10.1039/C4NR01600A}.
\newblock URL \url{http://dx.doi.org/10.1039/C4NR01600A}.
\newblock Publisher: The Royal Society of Chemistry.

\bibitem[Feurer et~al.(2015)Feurer, Klein, Eggensperger, Springenberg, Blum,
  and Hutter]{feurer_efficient_2015}
Matthias Feurer, Aaron Klein, Katharina Eggensperger, Jost Springenberg, Manuel
  Blum, and Frank Hutter.
\newblock Efficient and {Robust} {Automated} {Machine} {Learning}.
\newblock In \emph{Advances in {Neural} {Information} {Processing} {Systems}
  28}, pages 2944--2952. Curran Associates, Inc., 2015.

\bibitem[Jones et~al.(1998)Jones, Schonlau, and Welch]{jones_efficient_1998}
Donald~R. Jones, Matthias Schonlau, and William~J. Welch.
\newblock Efficient {Global} {Optimization} of {Expensive} {Black}-{Box}
  {Functions}.
\newblock \emph{J. of Global Optimization}, 13\penalty0 (4):\penalty0 455--492,
  December 1998.
\newblock ISSN 0925-5001.
\newblock \doi{10.1023/A:1008306431147}.
\newblock URL \url{http://dx.doi.org/10.1023/A:1008306431147}.

\bibitem[Kotthoff et~al.(2017)Kotthoff, Thornton, Hoos, Hutter, and
  Leyton-Brown]{kotthoff_auto-weka_2017}
Lars Kotthoff, Chris Thornton, Holger~H. Hoos, Frank Hutter, and Kevin
  Leyton-Brown.
\newblock Auto-{WEKA} 2.0: {Automatic} model selection and hyperparameter
  optimization in {WEKA}.
\newblock \emph{Journal of Machine Learning Research}, 18\penalty0
  (25):\penalty0 1--5, 2017.

\bibitem[Lang et~al.(2019)Lang, Binder, Richter, Schratz, Pfisterer, Coors, Au,
  Casalicchio, Kotthoff, and Bischl]{lang_mlr3_2019}
Michel Lang, Martin Binder, Jakob Richter, Patrick Schratz, Florian Pfisterer,
  Stefan Coors, Quay Au, Giuseppe Casalicchio, Lars Kotthoff, and Bernd Bischl.
\newblock mlr3: {A} modern object-oriented machine learning framework in {R}.
\newblock \emph{Journal of Open Source Software}, 4\penalty0 (44), 2019.
\newblock URL \url{https://doi.org/10.21105/joss.01903}.
\newblock 1903.

\bibitem[Lin et~al.(2014)Lin, Peng, Liu, Ruiz-Zepeda, Ye, Samuel, Yacaman,
  Yakobson, and Tour]{lin_laser-induced_2014}
Jian Lin, Zhiwei Peng, Yuanyue Liu, Francisco Ruiz-Zepeda, Ruquan Ye, Errol
  L.~G. Samuel, Miguel~Jose Yacaman, Boris~I. Yakobson, and James~M. Tour.
\newblock Laser-induced porous graphene films from commercial polymers.
\newblock \emph{Nature Communications}, 5\penalty0 (1):\penalty0 5714, December
  2014.
\newblock ISSN 2041-1723.
\newblock \doi{10.1038/ncomms6714}.
\newblock URL \url{https://doi.org/10.1038/ncomms6714}.

\bibitem[Luo et~al.(2016)Luo, Hoang, and Liu]{luo_direct_2016}
Sida Luo, Phong~Tran Hoang, and Tao Liu.
\newblock Direct laser writing for creating porous graphitic structures and
  their use for flexible and highly sensitive sensor and sensor arrays.
\newblock \emph{Carbon}, 96:\penalty0 522--531, 2016.
\newblock ISSN 0008-6223.
\newblock \doi{https://doi.org/10.1016/j.carbon.2015.09.076}.
\newblock URL
  \url{http://www.sciencedirect.com/science/article/pii/S0008622315302943}.

\bibitem[Marcano et~al.(2010)Marcano, Kosynkin, Berlin, Sinitskii, Sun,
  Slesarev, Alemany, Lu, and Tour]{hummers}
{Daniela C.} Marcano, {Dmitry V.} Kosynkin, {Jacob M.} Berlin, Alexander
  Sinitskii, Zhengzong Sun, Alexander Slesarev, {Lawrence B.} Alemany, Wei Lu,
  and {James M.} Tour.
\newblock Improved synthesis of graphene oxide.
\newblock \emph{ACS Nano}, 4\penalty0 (8):\penalty0 4806--4814, August 2010.
\newblock ISSN 1936-0851.
\newblock \doi{10.1021/nn1006368}.

\bibitem[Sygletou et~al.(2016)Sygletou, Tzourmpakis, Petridis, Konios, Fotakis,
  Kymakis, and Stratakis]{sygletou_laser_2016}
Maria Sygletou, Pavlos Tzourmpakis, Costas Petridis, Dimitrios Konios, Costas
  Fotakis, Emmanuel Kymakis, and Emmanuel Stratakis.
\newblock Laser induced nucleation of plasmonic nanoparticles on
  two-dimensional nanosheets for organic photovoltaics.
\newblock \emph{J. Mater. Chem. A}, 4\penalty0 (3):\penalty0 1020--1027, 2016.
\newblock \doi{10.1039/C5TA09199C}.
\newblock URL \url{http://dx.doi.org/10.1039/C5TA09199C}.

\bibitem[Wahab et~al.(2020)Wahab, Jain, Tyrrell, Seas, Kotthoff, and
  Johnson]{wahab_machine-learning-assisted_2020}
Hud Wahab, Vivek Jain, Alexander~Scott Tyrrell, Michael~Alan Seas, Lars
  Kotthoff, and Patrick~Alfred Johnson.
\newblock Machine-learning-assisted fabrication: {Bayesian} optimization of
  laser-induced graphene patterning using in-situ {Raman} analysis.
\newblock \emph{Carbon}, 167:\penalty0 609--619, 2020.
\newblock ISSN 0008-6223.
\newblock \doi{https://doi.org/10.1016/j.carbon.2020.05.087}.
\newblock URL
  \url{http://www.sciencedirect.com/science/article/pii/S0008622320305285}.

\bibitem[Wan et~al.(2018)Wan, Streed, Lobino, Wang, Sang, Cole, Thiel, and
  Li]{zhengfen_laser_2018}
Zhengfen Wan, Erik~W. Streed, Mirko Lobino, Shujun Wang, Robert~T. Sang,
  Ivan~S. Cole, David~V. Thiel, and Qin Li.
\newblock Laser-reduced graphene: Synthesis, properties, and applications.
\newblock \emph{Advanced Materials Technologies}, 3\penalty0 (4):\penalty0
  1700315, 2018.
\newblock \doi{https://doi.org/10.1002/admt.201700315}.

\bibitem[Wan et~al.(2019)Wan, Wang, Haylock, Kaur, Tanner, Thiel, Sang, Cole,
  Li, Lobino, and Li]{wan_tuning_2019}
Zhengfen Wan, Shujun Wang, Ben Haylock, Jasreet Kaur, Philip Tanner, David
  Thiel, Robert Sang, Ivan~S. Cole, Xiangping Li, Mirko Lobino, and Qin Li.
\newblock Tuning the sub-processes in laser reduction of graphene oxide by
  adjusting the power and scanning speed of laser.
\newblock \emph{Carbon}, 141:\penalty0 83--91, 2019.
\newblock ISSN 0008-6223.
\newblock \doi{https://doi.org/10.1016/j.carbon.2018.09.030}.
\newblock URL
  \url{http://www.sciencedirect.com/science/article/pii/S0008622318308443}.

\bibitem[Wang et~al.(2018)Wang, Wang, Zheng, Dong, Mei, Lv, Duan, and
  Wang]{wang_laser-induced_2018}
Fangcheng Wang, Kedian Wang, Buxiang Zheng, Xia Dong, Xuesong Mei, Jing Lv,
  Wenqiang Duan, and Wenjun Wang.
\newblock Laser-induced graphene: preparation, functionalization and
  applications.
\newblock \emph{Materials Technology}, 33\penalty0 (5):\penalty0 340--356,
  2018.
\newblock \doi{10.1080/10667857.2018.1447265}.
\newblock URL \url{https://doi.org/10.1080/10667857.2018.1447265}.

\bibitem[Ye et~al.(2015)Ye, Peng, Wang, Xu, Zhang, Li, Nilewski, Lin, and
  Tour]{ye_situ_2015}
Ruquan Ye, Zhiwei Peng, Tuo Wang, Yunong Xu, Jibo Zhang, Yilun Li, Lizanne~G.
  Nilewski, Jian Lin, and James~M. Tour.
\newblock In {Situ} {Formation} of {Metal} {Oxide} {Nanocrystals} {Embedded} in
  {Laser}-{Induced} {Graphene}.
\newblock \emph{ACS Nano}, 9\penalty0 (9):\penalty0 9244--9251, September 2015.
\newblock ISSN 1936-0851.
\newblock \doi{10.1021/acsnano.5b04138}.
\newblock URL \url{https://doi.org/10.1021/acsnano.5b04138}.
\newblock Publisher: American Chemical Society.

\end{thebibliography}

\end{document}